\documentclass[twocolumn, compsoc,9pt]{IEEEtran}
\usepackage{preamble}  
\usepackage{textcomp}
\usepackage{stmaryrd}
\usepackage[super,sort]{cite}
\makeatletter \renewcommand{\@citess}[1]{\textsuperscript{[#1]}} \makeatother
\makeatletter
\makeatletter
\renewcommand\section{\@startsection {section}{1}{\z@}%
                                   {-.5ex \@plus -1ex \@minus -.2ex}%
                                   {.2ex \@plus.2ex}%
                                   {\color{black}\Large\bf\sffamily}}
\renewcommand\subsection{\@startsection {section}{1}{\z@}%
                                   {-.5ex \@plus -1ex \@minus -.2ex}%
                                   {0.5ex \@plus.2ex}%
                                   {\color{darkgray}\Large\bf\sffamily}}
\makeatother
\begin{document}  
\def\TITLE{\vskip .5em \Huge   Algorithmic    Bio-surveillance For  Precise  Spatio-temporal Prediction of  Zoonotic Emergence }
\twocolumn[
\xtitaut{\fontsize{24}{24}\selectfont\TITLE}{\sffamily \fontsize{8}{12}\selectfont
Jaideep Dhanoa$^{\dag}$ \\
Balaji  Manicassamy$^{\mathsection}$ \\
Ishanu Chattopadhyay$^{\dag\ddag\star}$ \\ \vskip .2em
\rm \sffamily  \fontsize{8}{8}\selectfont
$^{\dag}$Department of Medicine, University of Chicago, Chicago IL, USA\\
$^{\mathsection}$Department of Microbiology, University of Chicago, Chicago IL, USA\\
$^{\ddag}$Institute for Genomics \& Systems Biology, University of Chicago, Chicago IL, USA\\
$^{\star}$Corresponding Author. \texttt{(ishanu@uchicago.edu)}
}
]

%\vspace{5pt}

{\begin{abstract} 
%\bf\sffamily 
\bf \fontsize{8}{10}\selectfont 
%\noindent \color{black!90} 
Viral zoonoses have emerged as the key drivers of  recent pandemics. Human infection  by zoonotic viruses  are either spillover events -- isolated  infections that fail to cause a widespread contagion -- or species jumps, where  successful adaptation to the new host leads to a pandemic. Despite expensive bio-surveillance efforts,  historically emergence response has  been reactive, and post-hoc. Here we use machine  inference to demonstrate a high accuracy  predictive bio-surveillance capability, designed to pro-actively localize an impending species jump via automated interrogation of massive sequence databases of viral proteins. Our results suggest  that a  jump might  not purely be the result of an isolated unfortunate cross-infection localized in space and time; there are subtle yet detectable patterns of genotypic changes accumulating in the global viral population leading up to emergence. Using tens of thosands of  protein sequences simultaneously, we train models that track  maximum achievable accuracy for disambiguating host tropism from the primary structure of  surface proteins, and show  that the inverse classification accuracy is a  quantitative indicator of  jump risk. We validate our claim in the context of the 2009 swine flu outbreak, and the 2004 emergence of H5N1 subspecies of Influenza A from avian reservoirs; illustrating  that interrogation of the global viral population  can unambiguously track a near monotonic risk elevation over several preceding years leading  to  eventual emergence. 
%This study lays the groundwork for an effective, feasible, and relatively cheap approach to persistent predictive global bio-surveillance.
% Most emerging human diseases are zoonoses, which are infections caused by pathogens of
% animal origin (Taylor et al., 2001). Early detection of potentially high-risk pathogens within
% animal hosts or vectors could enable mitigation strategies to prevent a species jump to
% humans, such as avoidance of high-risk areas, prophylactic drug distribution or timely
% mobilization of surveillance and medical resources to cope with emergent disease. However,
% our understanding of host–pathogen ecology and evolution is not yet sufficiently robust to
% allow us to recognize the patterns, processes and mechanisms that predicate species jumps.
% In the future, persistent surveillance in animals could detect changes in viruses that precede
% a species jump and allow mitigation or prevention of human infections. The prospects for
% predicting infectious disease outbreaks have been reviewed and discussed by several authors
% (Cleaveland et al., 2001; Taylor et al., 2001; Childs, 2004; Wolfe et al., 2005; Holmes and
% Drummond, 2007; Parrish et al., 2008; Childs and Gordon, 2009; Pulliam and Dushoff,
% 2009; Pepin et al., 2010). In this review, we outline a conceptual framework for achieving a
% virus surveillance capability that could predict future species jumps.
\end{abstract}}
\begin{IEEEkeywords}
bio-surveillance, Influenza A, antigenic shift, pandemic
\end{IEEEkeywords}
\tikzexternalenable 
\vspace{18pt}    

\allowdisplaybreaks{
%\section*{Background}
%

%\lettrine[lraise=0.1, nindent=0em, slope=-.5em]{E}{merging}
\IEEEPARstart{E}{merging} human diseases are often  infections caused by pathogens of
animal origin~\cite{pmid11516376,Flanagan2012} (zoonoses). Identification of  high-risk pathogens within
animal hosts can be  used to pro-actively trigger  mitigation strategies, potentially reducing the risk of  a successful  jump to
humans. However,
our incomplete understanding of host-pathogen interaction hinders preemptive recognition of subtle  signals that elevate the jump risk.
A complex interplay of the standing viral population, animal  and human hosts, environmental and socio-economic factors, make the task of identifying viruses of high zoonotic or pandemic risk, before emergence,  difficult to uncertain at best~\cite{pmid11516377,pmid16485465,pmid17848060,pmid18772285,pmid19787654,pmid19281304,pmid20938453,Flanagan2012}. 

Here we present an efficient,  data-driven  approach to persistent predictive bio-surveillance. At the core of our approach is an inference  algorithm to  
estimate    dissimilarity between  distinct viral populations, viewed as ensembles of protein sequences. In contrast to distance calculations in  phylogenetic analyses, where one computes a distance between two individual sequences~\cite{hannenhalli1995transforming,jean2007genome,ozery2003two,tesler2002efficient,shao2012approximating}, here we compute the  dissimilarity or distance  between two sequence ensembles. Unlike static distance formulae, our measure adapts to the evolving populations to back out the most important set of disambiguating residues (features) for the two populations. Computing, in this manner, the instantaneous dissimilarity between the host-specific viral quasi-species leads us to a time-varying  measure of jump risk. As an example, we claim that greater  the  similarity  between  the population of human influenza  viruses   and those currently prevalent in swines, higher  the possibility  of a species jump.

In machine learning parlance, our algorithm   trains a   classifier: given two sets of amino acid sequences for a specific viral protein corresponding to the two host species, it infers the optimal set of decision rules that  disambiguate the  populations with maximum achievable accuracy.    Then, dissimilarity  is simply the inverse  accuracy for the learned model. The interpretation here is  the tautology that ``similar'' objects are harder to distinguish, and hence lower classification accuracy indicates a higher degree of similarity. The inferred  classifier evolves with time, always distilling the optimal set of disambiguating rules to separate the populations. This adaptive tracking of the evolutionary changes, along with the elimination of  the choice of which static distance to use, provides us with  a more natural framework to discern subtle changes across viral populations.
%####################################################
%####################################################
%####################################################
%####################################################
\begin{figure*}[!ht]
\centering
\includegraphics[width=\textwidth]{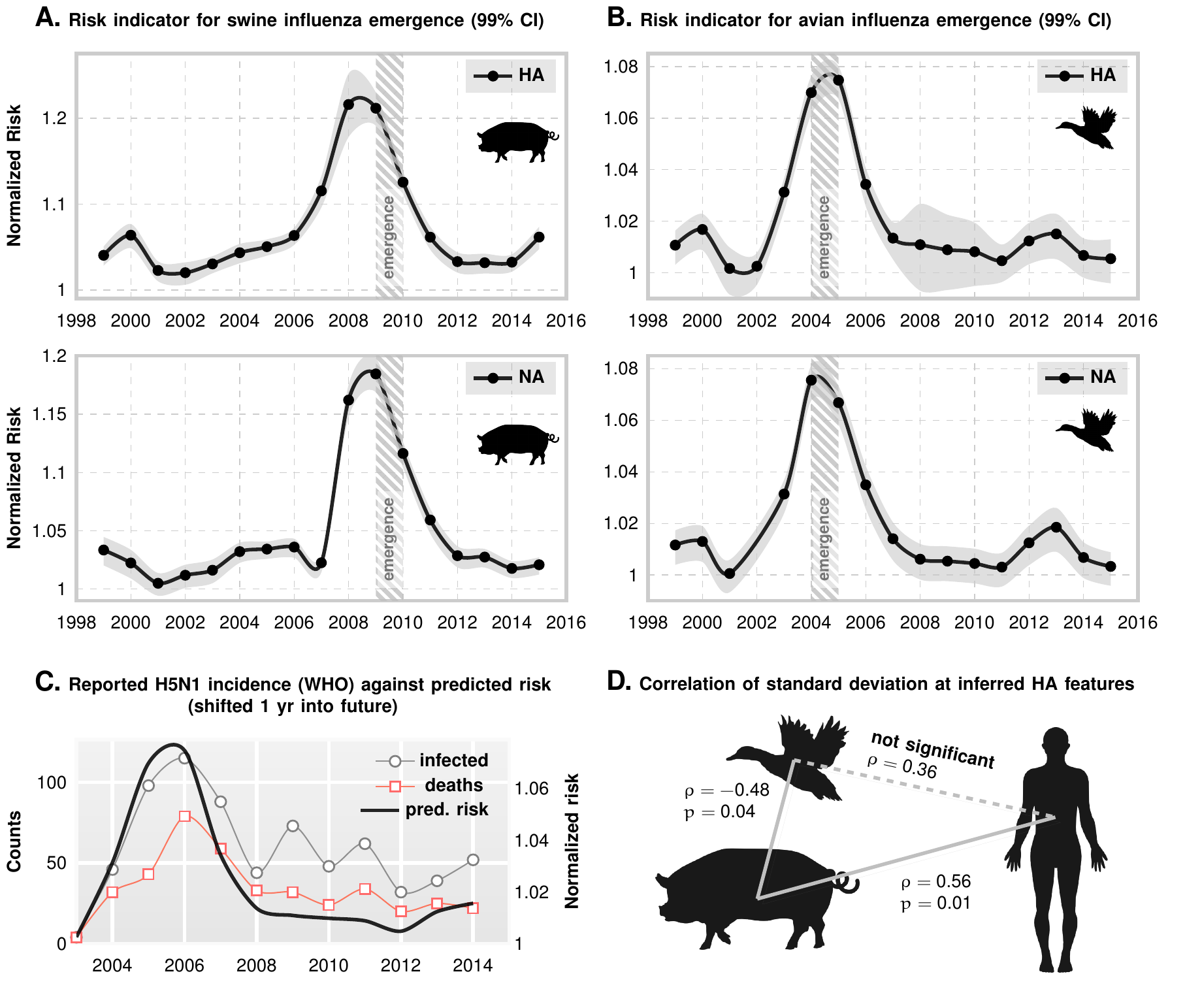}
\vspace{-20pt}

\captionN{\textbf{Main Results.} Automated inference  of emergent patterns in  host-specific HA and NA protein sequences (targeting human, swine and avian hosts) from the Influenza Research Database (IRD), distills an algorithmic risk predictor  for zoonotic emergence for influenza. Plates A and B illustrate  risk inference for the cases of the 2009 swine flu and the 2004 H5N1 emergence events. In both cases, we see a near monotonic risk elevation leading up to the event, with multiple years of actionable warning. \textit{Importantly, the inference algorithm only uses  past information at each predicted time-point.}  Except for small variations in accuracy, similar results are obtained for both HA and NA sequences. This is not surprising: while NA is not directly implicated in cellular entry, it is known to  assist  in transmission via enabling release of progeny viruses~\cite{Air1985}. Our algorithm  is not specific to influenza, and is applicable generally for predicting  zoonotic emergence. Plate C compares the predicted  H5N1 emergence risk (appropriately scaled)  to incidence reported by WHO~\cite{WHO}. We shifted the risk plot by 1 year into future to illustrate the close match, $i.e.$, our prediction closely pre-empted the  overall  incidence dynamics (positive correlation of 0.88 (with death counts) with p-value less than 0.0001). Plate D illustrates the correlation between residue specific standard deviations for host pairs as they evolve over time, where we use the same set of residues (See Table~\ref{tabresidue}) as identified by our algorithm to have  predictive value. We note that the swine-human and avian-human correlations are significant, while the avian-human is not; potentially corroborating the idea of domestic pigs as mixing vessels~\cite{Smith2009,Joseph2017} (See Discussion).
}\label{figrisk}
\end{figure*}
%####################################################
%####################################################
%####################################################
%####################################################
 
\subsection*{Key Insight}
With the application of the  inverse classification accuracy in  estimating  jump risk, we are putting  forward (and eventually validating) a key hypothesis: emergence risk   may be estimated accurately by looking for subtle sequence changes over time  in circulating strains.  Underlying conventional post-hoc reconstruction of emergence pathways, there is the assumption that species jumps are the  result of an unfortunate sequence of antigenic shifts |  abrupt genetic rearrangements between distinct strains co-infecting the same host cell, that dramatically alter the antigenic makeup of the resultant virus. Our hypothesis, if true, would imply that such reconstructions do not convey a complete picture of the processes and interactions that foster emergence. 

The  2009 pandemic strain (pH1N1) serves as a good example. The emergent strain became known as ``swine flu'', on account of pH1N1's  strong similarities with the then circulating swine influenza viruses; phylogenetic analyses showed that the pH1N1 genes clustered with those from swine viruses rather than the   seasonal human flu strains. Further analysis suggested that pH1N1 resulted from the  re-assortment of 2, or even 3, distinct viruses, namely the Eurasian swine  H1N1, and the swine H1N2; the latter itself having emerged from swine H1N1 and the  triple assortment swine strain trH3N2, which  in turn had contributions from the human H3N2 (related to the Hong Kong flu epidemic of 1968), and even had similarities to avian strains circulating in north America~\cite{VanderMeer2010,Smith2009}. 
It is generally recognized that such reconstructions of evolutionary pathways are not unique. Alternate event sequences might have transpired in practice, particularly since swine H1-containing viruses regularly spill-over to humans without causing widespread infections. Additionally, while all pH1N1 genes appear to have originated in swines, they come from  geographically widely distributed ancestors. One explanation to this ancestral diversity  is the possibility that pH1N1 emerged over a span  several years,  cryptically circulating in swines before pandemic recognition~\cite{Smith2009}. 
Irrespective of the specific details, if antigenic shifts are solely responsible for species jumps, then emergence is precipitated entirely by chance events; and hence is categorically impossible to predict |  even with vast surveillance efforts. In contrast, our  hypothesis suggests that  gradual processes, such as antigenic drift brought about by point mutations continuously altering the transcribed proteins over time, play a crucial role; in essence  setting up the stage for the re-assortment event that leads to emergence.

%A crucial consequence of our  results  is the practical ability to carry out truly predictive surveillance. 
Our  risk indicator does not require identification of the specific originating animal. Global sampling of the host-specific viral populations  suffices to track the progressive similarity of the  populations, and  a near monotonic risk elevation  leading up to the jump. If we need to somehow locate the specific animal(s) in which a new virus emerges  in time | every time | then,  it is 
ultimately a losing battle. For example, the 2009 pandemic strain was isolated in a specific pig farm months after the first reported human infections~\cite{VanderMeer2010}.  However, if we can reliably  estimate  jump risk in space, time and originating species  by merely 
sampling animals across the globe, and individual members of the  host species  are  less important, then we shift the odds in our favor.

%#########################################################
%#########################################################
%#########################################################
%#########################################################
%#########################################################
%#########################################################
\begin{figure}[!ht]
\includegraphics[width=0.45\textwidth]{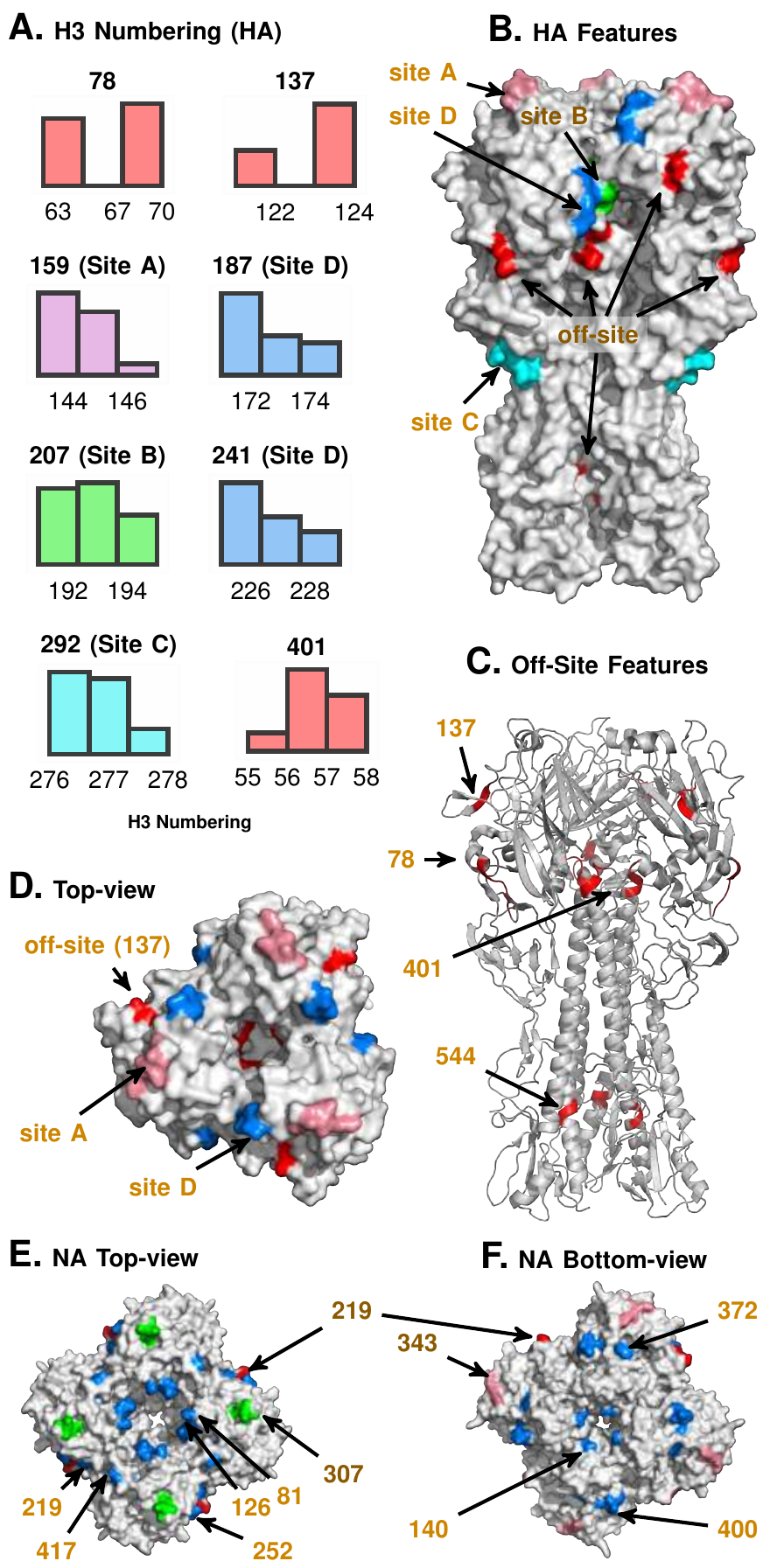}
\vspace{15pt}

\captionN{\textbf{Location of Inferred Residues of Predictive Value in 3D Molecular Structure.} As expected, a subset of the inferred residues  are close to known antigenic sites. For HA, the minimal list of such key features consist of $9$ residues, of which 5 correspond to the four classical antigenic sites A,B,C,D~\cite{Weis1988}, while the rest are not in regions that generally contribute to monoclonal antigenicity. We use sequential numbering for these residues, and since we analyze sequence ensembles, individual features map to a distribution in the  H3 numbering scheme (shown in Plate A). Plates B-D show  our inferred HA features, and plates E-F show the inferred features for NA. All of our inferred feature are not surface residues; features 400-402, and 544  for HA, and 22, 48, 51, 97, 182, 204 for NA are not exposed on the surface of the trimer and tetramer respectively (See Table~\ref{tabresidue} for complete list of inferred features). 
The appearance of these residues are surprising; but too predictive to be ignored.
}\label{figmol}

\vspace{-18pt}

\end{figure}
%##################################
%  %#############################
%  %#############################
%  %#############################
\begin{figure*}[t]
\centering 
 \includegraphics[width=0.8\textwidth]{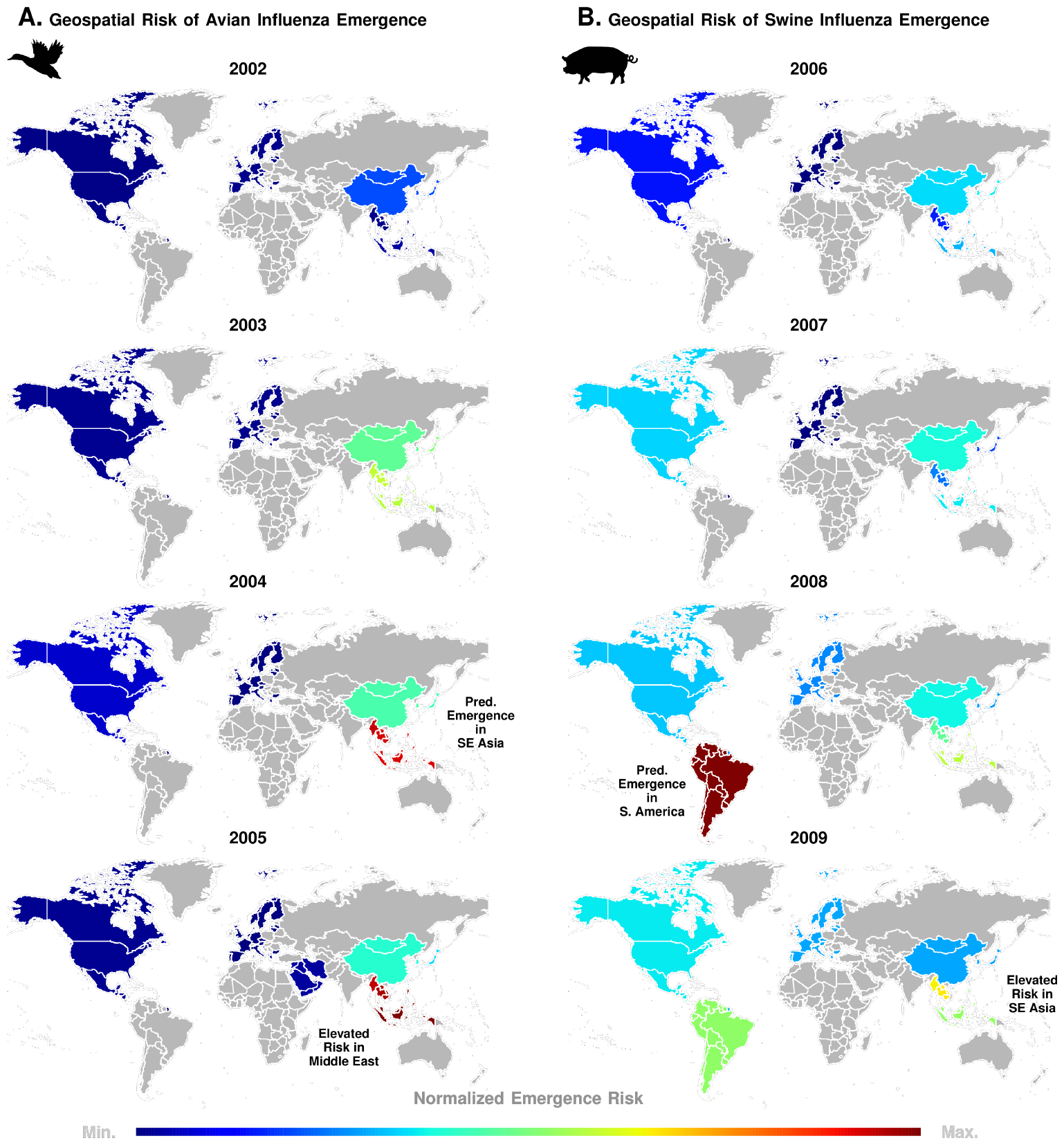}

\captionN{\textbf{Geo-spatial Emergence Prediction.} Our algorithm may be used to geographically localize the emergence risk, by feeding it geographically stratified sequence data. The key challenge is the sparsity of sequences from around the world in the IRD, which degrades our accuracy. Nevertheless, as shown in columns A and B, we correctly localize both the 2004 H5N1 and the 2009 swine flu emergence. Note that we could not predict the risk elevation in Mexico prior to 2009 due to the extreme sparsity of collected sequences for S. America. Additionally, the algorithm also predicts correctly the risk elevation in the middle east in 2005 for the avian flu emergence, and the SE Asia in 2009 immediately after the swine flu outbreak.  }\label{figworld}
\end{figure*}
%  %#############################
%  %#############################
%#############################
%############################# 
\begin{table*}[t]
\centering

\captionN{Classification problem  setup: Human \& Swine Influenza A Viruses (HA Sequences, standard code for amino acids)}\label{tabseq}
\def\WDTC{-0.1in}

\def\COLX{black!30}
\def\COLZ{black!25} 
\def\COLY{black!20}
\bf\sffamily\fontsize{5}{6}\selectfont
\begin{tabular}{ccccccccccccccccccccccc}
Species&275&276&277&278&279&280&281&282&283&284&285&286&287&288&289&290&291&292&293&294&295& $\large\cdots$\\\cellcolor{\COLX} &&&&&\\
\cellcolor{\COLX}Swine&S&R&G&L&G&S&G&I&I&T&S&K&A&P&M&D&E&C&D&A&K& $\large\cdots$\\
\cellcolor{\COLX}Swine&S&R&G&L&G&S&G&I&I&T&S&K&A&P&M&D&E&C&D&A&K& $\large\cdots$\\
\cellcolor{\COLX}Swine&F&K&I&R&R&G&K&S&S&I&M&R&S&D&A&P&I&G&K&C&N& $\large\cdots$\\
\cellcolor{\COLX}Swine&G&R&G&L&G&S&G&I&I&T&S&K&A&P&M&D&E&C&D&A&K& $\large\cdots$\\
\cellcolor{\COLX}Swine&G&R&G&L&G&S&G&I&I&T&S&K&A&P&M&D&E&C&D&A&K& $\large\cdots$\\
%\cellcolor{\COLZ} &\\
\cellcolor{\COLY}Human&F&K&I&R&S&G&K&S&S&I&M&R&S&D&A&P&I&G&K&C&K& $\large\cdots$\\
\cellcolor{\COLY}Human&M&E&R&N&A&G&S&G&I&I&I&S&D&T&P&V&H&D&C&N&T& $\large\cdots$\\
\cellcolor{\COLY}Human&M&E&R&N&A&G&S&G&I&I&I&S&D&T&P&V&H&D&C&N&T& $\large\cdots$\\
\cellcolor{\COLY}Human&M&E&R&N&A&G&S&G&I&I&I&S&D&T&P&V&H&D&C&N&T& $\large\cdots$\\
\cellcolor{\COLY}Human&M&E&R&N&A&G&S&G&I&I&I&S&D&T&P&V&H&D&C&N&T& $\large\cdots$
\end{tabular}
\centering
\vskip 3em
\captionN{Inferred Predictive Features (Features are numbered in the sequential scheme)}\label{tabresidue}
% \begin{tabular}{C{.5in}C{.6in}C{.6in}}
%     \hline
%     \multirow{2}{*}{Multirow}&X&\multirow{2}{*}{78,137,402}\\\cline{1}
%     &X\\
%     \hline
% \end{tabular}
\bf\sffamily\fontsize{7}{8}\selectfont
\def\CCOL{lightgray!40}
\def\CCOLA{Honeydew1}
\def\CCOLB{white}
\def\DCOL{lightgray!40}
\begin{tabular}{C{.5in}C{.8in}|C{.8in}|C{.8in}|C{.8in}|C{.8in}
C{1in}C{1.5in}}
    \hline
Protein&\multicolumn{5}{c}{\cellcolor{\CCOL}Inferred Predictive Residues (Features)}&Minimal Feature Set\\\hline
\cellcolor{\DCOL}HA& 157, 158, 159   & 205, 207, 208  & 290, 291, 292  & 240, 241, 242 & 77, 78, 137, 400, 401, 402,  545 &\cellcolor{\CCOL}\mnp{1in}{\vskip 1ex 78, 137, 157, 187\vskip 1ex 207,  291, 241, 401}\\\cline{2-6}
\multirow{-2}{*}{\cellcolor{\DCOL}}&\cellcolor{\CCOLA}site A~\cite{Weis1988}&\cellcolor{\CCOLA}site B~\cite{Weis1988}&\cellcolor{\CCOLA}site C~\cite{Weis1988}&\cellcolor{\CCOLA}site D~\cite{Weis1988}&offsite&\multirow{-2}{*}{\cellcolor{\CCOL}}\\\cline{1-7}
\\\hline
\cellcolor{\DCOL}NA&\multicolumn{3}{c|}{204, 215, 219, 252, 343, 346, 372, 400}&\multicolumn{2}{c}{ \mnp{1.9in}{\vskip 1ex22, 48, 51, 81, 84, 97, 126, 140, 141\vskip 1ex 182, 309, 307, 417\\}}&\cellcolor{\CCOL}48, 51, 97, 219, 307, 344\\\cline{2-6}
\multirow{-2}{*}{\cellcolor{\DCOL}}&\multicolumn{3}{c|}{\cellcolor{\CCOLA}close to antigenic sites~\cite{Saito1994,Air1985}}&\multicolumn{2}{c}{offsite}&\multirow{-2}{*}{\cellcolor{\CCOL}   }\\\cline{1-7}

\end{tabular}

\end{table*}
%#############################
%#############################
\subsection*{Quantifying Jump Risk For Influenza A}
\vspace{-6pt}

Influenza  is responsible for one of the most devastating 
epidemics in human history, decimating over 2\% of the human population in  the H1N1 Spanish flu outbreak of 1918-1920. In addition to be implicated in  tens of thousands of deaths every year in US alone  from the recurring seasonal flu epidemic, influenza continues to emerge  again and again in humans from strains circulating in animals, leading to severe to moderate spikes in incidence and mortality rates. Two such recent pandemics are the 2004 emergence  of the highly pathogenic H5N1  avian strain, and the  pH1N1 swine flu outbreak of 2009.
Given the fact that all known  influenza subtypes have been isolated in birds~\cite{pmid15709000,pmid1579108,pmid16627734}, and that all pandemics with the exception of the 2009 event were caused by strains of avian origin~\cite{VanderMeer2010}, surveilling avian strains is of paramount importance. With the emergence of pH1N1 with its complicated genetic ancestry causing between 151,700 and 575,400 deaths~\cite{Wang2013}, it is also imperative  that we closely monitor swines for future emergence. 
These recent events, along with the  availability of large  databases of influenza proteins (Influenza Research Database or IRD~\cite{IRD}),  prompted us to select avian and swine Influenza A viruses as validation candidates for our general  bio-surveillance algorithm. 

Influenza A is a negative stranded RNA virus % from the \textit{Orthomyxoviridae} family, 
with an encapsulated segmented genome surrounded by   the host cell-derived lipid membrane.
We focus on the two glycoproteins embedded in the envelope membrane,  hemagglutinin (HA) and neuraminidase (NA), implicated respectively in cellular entry and release of progeny viruses. Due to their surface exposure,  antigenicity of HA and NA  categorizes  influenza A viruses into 17 currently known subtypes of HA (H1 to H17) and ten of NA (N1 to N10). With  segmented genome facilitating  re-assortment with different strains,  the virus is able to emerge with a new suite of segments and subtypes~\cite{pmid12660783,Mair2014}. We hypothesized that the chances of these antigenic shifts  are modulated, and foreshadowed, by incipient  patterns in the sequences of the  circulating strains. And  that these patterns  may be  distilled from the IRD via appropriate statistical analyses.
%, which  may then be used to build an effective indicator of jump risk. 

%To  identify the statistical footprint of an impending jump, we set up an automated classification problem. 
Querying the IRD for  all relatively recent and complete HA and NA  sequences, we  ended up with 26,635, 7696 and 16,696 HA, and 22,488, 7662 and 14,205 NA sequences  for human, swine and avian hosts respectively, collected within the 17 year period between 1999 and 2016. The restriction to this time period arose from the necessity to have a minimum number of sequences each year for reliable statistical analysis.
With the objective of modeling  the differences between host-specific strains at any given point in time, we did not distinguish between antigenic  subtypes. We expected that our classification algorithm to automatically distinguish residue differences dictating sub-type categorization if necessary. 
Additionally, we used sequential numbering for referring to the residue positions, and did not attempt to globally align the collected sequences. Not using a standardized scheme (such as H3 numbering for HA, and N2 numbering for NA) is driven by the idea that for a large enough collection of sequences, the random variations at each sequential position (which would be reduced by aligning to a reference sequence in the standardized numbering process)  might  be key to unraveling important predictive patterns. 

A  small  excerpt of the HA sequences for human and swine influenza between residues 275 and 295 (sequential numbering) is shown in Table~\ref{tabseq}. For the majority of the  residues,  there are  variations  within each species, as well as across.  We asked  if, given a sufficiently large  set of  sequences  collected  within some relatively short period of time (1 year),  we can train a protein-specific  classifier that accurately models these subtle   patterns of variation  to reliably recognize the host species. We found that relatively  simple decision trees are able to adequately model the species specific patterns with high out-of-sample accuracy reaching  $95\%$-$99\%$ (See Fig.~\ref{figsupp1}, plates A-D). For example, a couple of  rules encoded by the decision tree shown in plate B of Fig.~\ref{figsupp1} are: \textit{if  residue  78 is I or K, and  residue 292 is N, D or T, then the HA sequence is from a human host with less than 1\% error.} On the other hand, \textit{if the residue  78 is I or K, and the residue  292 is K or E, and the residue  400 is V, then the host is swine with approximately 6\% probability of error.} 
The tree encodes $5$ such rules in total, each of which terminates in a distinct leaf of the tree (the nodes at the bottom layer). 
The  structure of the inferred tree  corresponds to the number and complexity of the encoded decision rules, which vary with the time period of collection of the sequences, the host species involved, and the protein under study.

These  decision trees are computed using unbiased recursive partitioning~\cite{Hothorn2006} on sets of host-specific sequences drawn within a period of 1 year. We measure model performance  on the training data with  \textit{in-sample accuracy}: which is the fraction of correct classifications on the training data itself once the model is inferred. We also test   performance on data not used during training, $i.e.$, sets of sequences not from the same time period within which a particular classifier is trained, by computing the \textit{out-of-sample accuracy}.

The key computational challenge here arises from the existence of  many possible alternate choices of decision rule-sets that disambiguate the host species. This  redundancy partially arises from  dependencies among non-colocated residues required for correct assembly and  function~\cite{Myers2013}. Here we aim to curate the minimal set of residues that disambiguate the hosts (irrespective of the time period), and such dependencies imply that  numerous equally accurate sets of rules exist. We solve this issue via \textit{iterative feature depletion}: we construct a conditional inference tree, identify the \textit{most important residue} (one that has maximum contribution in classification accuracy), delete that feature from the training algorithm, and re-run the tree inference. As we continue to iterate in this manner, in each step we 
compute the out-of-sample accuracy by applying the learned model on sequences from all other one year time periods. We stop if the out-of-sample accuracy falls below 90\%, or if we run out of features. Carrying out this iterated deletion for all time-periods, we identify a sequence of decision trees, all of which are highly accurate models of host tropism, irrespective of the time period of analysis. Charting the number of times each residue appears as the most important feature, we end up with a small set  that have maximal contribution in recognizing the target host. Once this set is identified, we train a \textit{random forest classifier~\cite{Breiman2001}} with the residues as features, for each year. The in-sample accuracy achieved by these forests are  then inverted to compute the year-specific jump risk. Our  results  for HA and NA, and for swine-human and avian-human jumps is shown in Fig.~\ref{figrisk} plates A-B.  The overall workflow of our algorithm is summarized in Fig.~\ref{figsupp1} plate E.

We can also restrict our algorithm to only access sequence data collected from just one country at a time,  to construct a geospatial estimate of the time-varying jump risk (See  Fig.~\ref{figworld}). Due to the severe sparsity of sequences in the IRD for many countries (See Fig.~\ref{figSI0} plate E), our geospatial predictions are relatively patchy, incomplete and suffers from widened confidence intervals. Nevertheless, we are able to pinpoint  correctly the time and place  of both the 2004 and 2009  events.

%  %#############################
%  %#############################
% %############################
% %############################
\begin{figure*}[!ht]
\includegraphics[width=\textwidth]{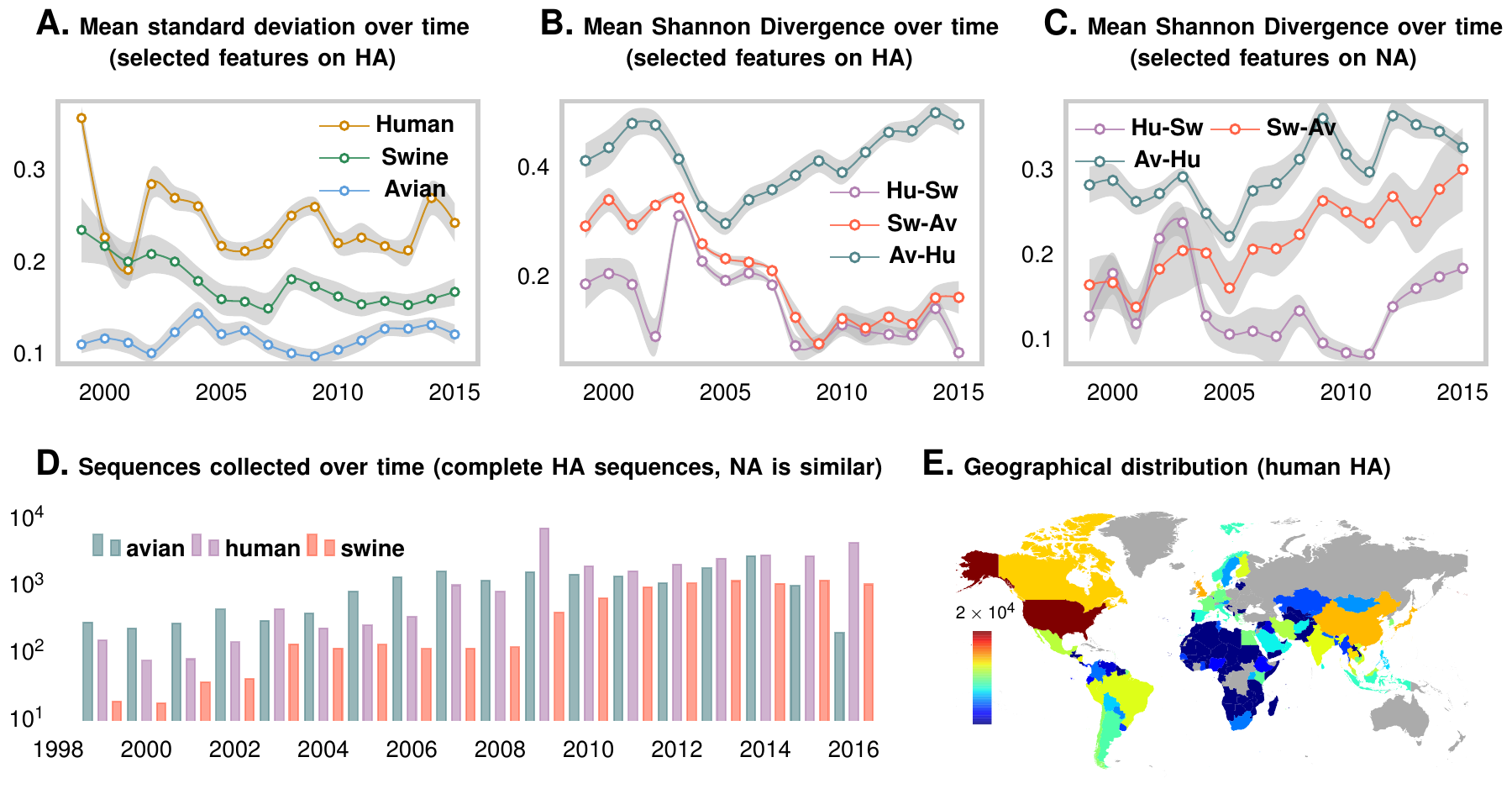}
\vspace{-15pt}

\captionN{Given the minimal set of predictive features identified by our algorithm, we computed the   variance at these residues for the host-specific strain ensembles, as the virus continues to evolve. As shown in plate A, we get a strong  and significant  positive correlation between human and swine specific strains, and a significant strongly negative correlation between avian and swine specific strains. the correlation between human and avian strains was also strongly negative, but not significant. Plates B and C show the mean Shannon divergence at the identified features for each pair of hosts. We see that for HA, the distance between human-swine and swine-avian roughly remains constant, whereas  the distance between the swine and  avian strains continues to diverge. Plate D shows number of sequences collected in the IRD over time, and plate E illustrates the geospatial imbalance in the database. The imbalance is more severe for swine and avian sequences. Importantly, we control for this imbalance, and we do not predict risk spikes only for places or times with most sequences. }\label{figSI0}
\end{figure*}
% %############################
% %############################
% %############################
% %############################
%############################
%############################
%############################
%############################
%\renewcommand{\figurename}{Supplementary Fig.}
%\setcounter{figure}{0}  
\begin {figure*}
\centering
\vspace{-17pt}

\includegraphics[width=\textwidth]{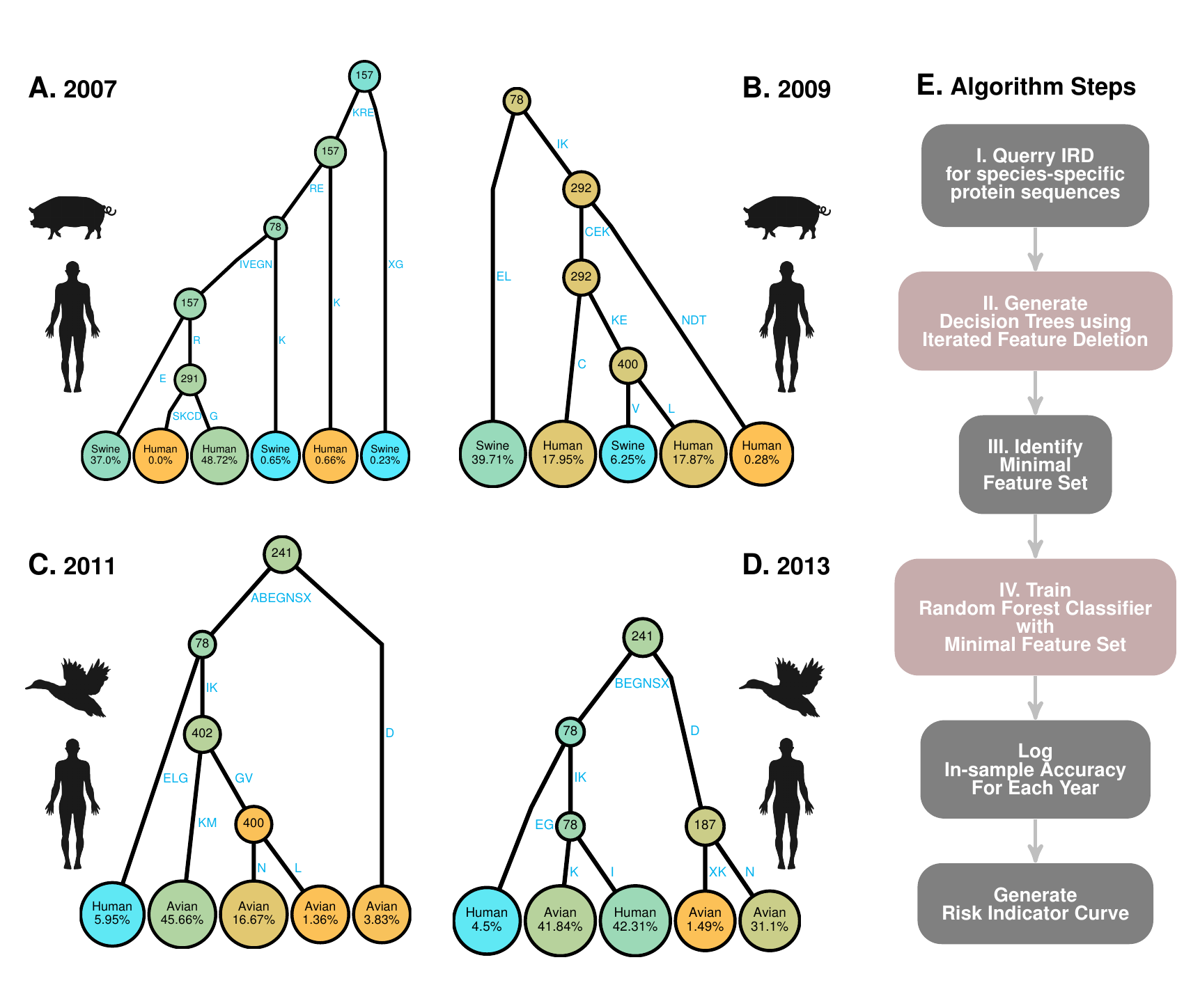}

\vspace{-12pt}

\captionN{\textbf{Examples of Inferred Conditional Inference Trees. } Plates A-D illustrate conditional inference trees that recognize HA sequences pertaining to human vs swine (A,B) and human vs avian (C,D)  for the respective years 2207,2009,2011, and 2013. The leaf nodes enumerate the majority class, along with the percentage class error. The colors of the node depict the relative mixture of the host species. The numbers in the non-leaf nodes denote the residue index (sequential numbering). These decision trees characterize the optimally inferred rules that allow one to decide the host species given the amino acid sequence. Note that the number of rules vary from tree to tree and over the years. The in-sample accuracy of these classifiers is over 93\%, with out-sample accuracy greater than 90\% for immediate future. Plate E enumerates a summarized sketch of the algorithm, along with the key steps. Steps II and IV are the computational bottlenecks.  }\label{figsupp1}
\vspace{-5pt}

\end{figure*}
%############################
%############################
%############################
%############################
%############################

\subsection*{Discussion}
To summarize our computational approach, we construct viral host recognizers (for human, swine and avian Influenza A) by using the 
primary structure of HA and NA proteins, to first identify  a minimal set of residues that allow for good out-of-sample classification performance across the years, and then  using this invariant minimal feature set to estimate the maximum  in-sample classification accuracy for individual years. Finally, we  interpret this time-varying accuracy  as the inverse jump risk indicator for selected host-pairs.

Viral  populations evolve continuously; thus an invariant minimal set of  residues that disambiguate target hosts  reflect the seats of fundamental differences in molecular structures driving host-specific infection and transmission processes.
A known causal factor  is the specificity of HA binding to avian-like $\alpha$-2,3-sialic acid (SA)  versus the  mamalian-like $\alpha$-2,6-SA receptors~\cite{Li2013,Joseph2017}. Therefore, substitutions in and around the HA Receptor Binding Site (RBS)  possibly could  drive host specificity, and the HA minimal residue set we identified is consistent with this observation.

Structurally, the native HA is trimeric, and each monomer is comprised  of a distal domain of globular shape (HA1), and a proximal  stem  anchoring  into the viral lipid envelope (HA2)~\cite{Caton1982,Weis1988}. It is well-recognized that antigenic drift is driven by the  accumulation of amino acid substitutions in HA epitopes that block SA interaction~\cite{pmid6202836,pmid16925526}. The antigenic sites recognized by monoclonal antibodies with high neutralizing activity, tend to be  similar across subtypes~\cite{Caton1982}, and are generally categorized into  4 groups (A, B, C, D for H3, and Sa, Sb, Ca, Cb for H1 subtype~\cite{Caton1982,Hensley}). This number can change based on the specific sub-type~\cite{Caton1982}. Nevertheless, the residues we identified for HA have footprints in  all four  sites. Namely, in H3 numbering, the inferred  HA minimal feature set consists of residues  144 - 146 (site A, sequential index 159), 172 - 174 and 226 - 228 (site D, sequential indices 187 and 241), 192-194 (site B, sequential index 207), and 276-278 (site C, sequential index 292). In addition to residues within the antigenic sites, 4  other features appear in the minimal set: residues 63-70 (sequential index 78), 122-124 (sequential index 137), HA2 residue 55-58 (sequential index 401) and sequential index 544 near the lower end of the HA2 stem. The locations of these residues is shown in Fig.~\ref{figmol} plates B-D.  The occurrence of residues outside the RBS is not surprising, as such  mutations have been shown to be determinants of receptor binding specificity~\cite{Jayaraman2012,Imai2012}. 

Interestingly, not all residues in the minimal set have surface exposure. Nevertheless, these residues have been identified to have important roles in 
host specificity. HA-mediated membrane fusion in acidic environment is necessary for cellular entry~\cite{pmid11473264}, and  human viruses appear to fuse at a lower pH than avian and swine counterparts~\cite{pmid23486663,pmid25673693,pmid25653452,pmid23459660,pmid3089198,pmid4060851}. The residue in HA2 corresponding to sequential index 401 is near the tip of the fusion peptide, and substitutions in this region have been observed in experiments designed to characterize membrane fusion activity and virus stability~\cite{Baumann}. Substitutions in the  second HA2 residue at sequential index 544 has also being implicated in maintenance of thermal stability~\cite{Xu2013}, and proper expression of HA in cells.

Our second protein of interest, NA  is a homotetramer with each monomer consisting of a hydrophobic membrane anchor, a stalk, and a head region with the catalytic and antigenic domains~\cite{Air1985}. NA  cleaves SA receptors of host cells to enable dissemination of progeny viruses~\cite{webster2013textbook}, and an optimal balance between the HA and NA function is crucial: excess NA  hinders binding of HA to host cell receptors, whereas insufficient NA function limits  viral spread~\cite{pmid11987141,pmid21825167}. Similar to HA, NA has preferential specificity for $\alpha$-2,3-SA receptors in avian, and $\alpha$-2,6-SA receptors in mamalian viruses~\cite{pmid24668228}. As such, the  feature set for NA has major footprints within its known antigenic sites~\cite{Saito1994}. Of the initial set of residues identified, those at 204, 215, 219, 252, 343, 346, 372, 400 are near or at antigenic sites, whereas those at 22, 48, 51, 81, 84, 97, 126, 140, 141
182, 309, 307, 417 are not. Pruning these residues to a minimal set such that predictive performance is unaltered, we get a  set consisting of just 6 residues: 48, 51, 97, 219, 307, 344. Of these, 219 and 344 are on antigenic sites. Additionally,  97 is not exposed on the surface, and 48, 51 are not even on the head region. While the appearance of these later residues in the minimal feature set might be surprising, they have  significant contributions in prediction accuracy. (See Fig.~\ref{figmol} plates E and F).

The time-varying risk shown in Fig.~\ref{figrisk} plates A-B illustrate that an impending jump can be predicted years in advance from observing the ever increasing risk elevation. The avian risk indicator compares favorably, with appropriate scaling,  against the WHO report on H5N1 incidence since 2003 (See Fig.~\ref{figrisk}, plate C). While, we do not make a direct case that jump risk should translate to incidence rate, this close match is noteworthy.

We interpret these results  suggest that the viral populations circulating in the respective hosts are continuously interacting, and driving each other's molecular evolution. Without such  continuous interaction, it is difficult to see how one would get a gradual increase instead of the risk spiking just before emergence.  To investigate this claim further, we computed the mean standard deviation at the residues in the minimal feature set (for HA) over time (See Fig.~\ref{figrisk} plate D and Fig.~\ref{figSI0} plate A). The results show that with respect to this measure the human and swine strains are  strongly and significantly positively correlated ($\rho=0.56,p=0.01$), and the swine and avian and avian strains are strongly and significantly negatively correlated ($\rho=-0.48,p=0.04$). The negative correlation between the human and avian strains, on the other hand, is not statistically significant. While not conclusive, these results are consistent with the suggestion that domestic pigs act as mixing vessels~\cite{Smith2009,Joseph2017}. Additionally, these strong correlations also support the thesis that the circulating strains interact continuously, and drive antigenic change. 

We also computed the time-varying distance between host-pairs, measured as the average Shannon divergence at the residues of the minimal sets (for HA, in Fig.~\ref{figSI0}, plate B and for NA in Fig.~\ref{figSI0}, plate C). This distance for HA shows an intriguing pattern, it appears that the swine strains are equidistant on average from human and avian strains post 2004, whereas the avian human distance is increasing. The results are shown with 99\% confidence intervals. We hope that these results would spark new directions of research into the interaction dynamics of the host-specific strains.

In summary, the principal contribution of this study is an algorithmic approach to surveillance that exploits  subtle patterns of sequence changes. These results fundamentally challenge how we think about bio-surveillance: we do not need to seek out the individual animals in which a chance re-assortment event gives rise to a pandemic strain, we can carry out random sampling of the host species globally and still construct an accurate spatio-temporal picture of jump risk. While this study focuses on Influenza A in human, swine and avian hosts, the basic principles are expected to hold elsewhere: for other host species, and other zoonotic pathogens.

%############################

}%%%%% ALLOWDB

%\clearpage
\footnotesize
 
%\nocite{*}
\bibliographystyle{naturemag}
\bibliography{bioshock,bioshock_refs}
\end{document}